\documentclass[twocolumn,trackchanges]{aastex61}
\usepackage{graphicx}
\usepackage{epsfig}
\usepackage{rotating}

\received{\today}
\begin{document}

\title{X-ray census of Millisecond Pulsars in the Galactic field}
\correspondingauthor{David C. Y. Hui}
\email{cyhui@cnu.ac.kr, huichungyue@gmail.com}

\author{Jongsu Lee}
\affil{Department of Astronomy, Space Science and Geology, Chungnam
National University, Daejeon 34134, Korea}
\author{C. Y. Hui}
\affil{Department of Astronomy and Space Science, Chungnam
National University, Daejeon 34134, Korea}

\author{J. Takata}
\affil{Institute of Particle physics and Astronomy, Huazhong University of Science and Technology, China}

\author{A. K. H. Kong}
\affil{Institute of Astronomy, National Tsing Hua University, Hsinchu, Taiwan}

\author{P. H. T. Tam}
\affil{School of Physics and Astronomy, Sun Yat-sen University, Zhuhai 519082, China}

\author{K. S. Cheng}
\affil{Department of Physics, University of Hong Kong, Pokfulam Road, Hong Kong}

\begin{abstract}
We have conducted a systematic survey for the X-ray properties of millisecond pulsars (MSPs). Currently,
there are 47 MSPs with confirmed X-ray detections. We have also placed the upper limits for
the X-ray emission from the other 36 MSPs by using the archival data. We have normalized their X-ray luminosities $L_{x}$
and their effective photon indices $\Gamma$ into a homogeneous data set, which enable us to carry out a detailed
statistical analysis. Based on our censored sample, we report a relation of 
$L_{x}\simeq10^{31.05}\left(\dot{E}/10^{35}\right)^{1.31}$~erg/s (2-10~keV) for the MSPs. The inferred X-ray conversion efficiency is 
found to be lower than previously reported estimate that could be affected by selection bias. 
$L_{x}$ also correlates/anti-correlates with the magnetic field strength at the light cylinder $B_{LC}$/characteristic age $\tau$. 
On the other hand, there is no correlation between $L_{x}$ and their surface magnetic field 
strength $B_{s}$. 
We have further divided the sample into four classes: (i) black-widows, (ii) redbacks,
(iii) isolated MSPs and (iv) other MSP binaries, and compare the properties among them. We noted that while the rotational parameters
and the orbital periods of redbacks and black-widow are similar, $L_{x}$ of redbacks are significantly higher than those of black-widows
in the 2-10~keV band. Also the $\Gamma$ of redbacks are apparently smaller than those of black-widows, which indicates the X-ray emission of redbacks are
harder than that of black-widows. This can be explained by the different contribution of intrabinary shocks in the X-ray emission
of these two classes.
\end{abstract}

\section{Introduction}
Since the discovery of the first millisecond pulsar (MSP) PSR~B1937+21 (Backer et al. 1982), a distinct class
in the pulsar population, characterised by a rotational period $P\lesssim20$~ms and a spin-down rate $\dot{P}\lesssim10^{-18}$~s~s$^{-1}$,
has been established (Manchester et al. 2005). It is generally accepted that MSPs are formed when an old neutron star has been spun up
through  accreting mass and angular momentum from its companion (Alpar et al. 1982; Radhakrishnan \& Srinivasan 1982; Fabian et al. 1983).

In the recent years, the population of MSPs have been expanded significantly (e.g. Hui 2014 for a recent review).
Multi-wavelength follow-up investigations of unidentified $\gamma-$ray objects detected by {\it Fermi} Gamma-ray
Space Telescope have been demonstrated to be successful in discovering MSPs
(Abdo et al. 2013; Hui et al. 2015; Ray et al. 2012). Currently, there are $\sim200$ MSPs 
found in the Galactic field (Manchester et al. 2005).

Apart from expanding the population, observations in the recent years have also shown that MSPs can be further divided into several
sub-classes, including black-widows (referred as BWs hereafter), redbacks (referred as RBs hereafter) as well as isolated MSPs.

BWs are binary MSPs bounded in tight orbits ($P_{b}\lesssim20$~hrs) with companions of only a
few percents of solar mass (cf. Hui 2014). A prototypical example of this class is PSR~B1957+20 which is a binary system with a
9.2~hrs orbit containing a MSP with $P=1.6$~ms and a companion of $M_{c}\sim0.02M_{\odot}$ (Fruchter et al. 1988).
Another characteristics of the BWs is the presence of radio eclipse. In the case of PSR~B1957+20, the eclipses of the radio pulsations
occur regularly for $\sim10\%$ of its orbit (Fruchter et al. 1988).

RBs form a relatively new class of MSPs which has only emerged for a decade. Their orbital periods $P_{b}$ span a somewhat larger range than 
that of BWs (i.e. $P_{b}\lesssim20$~hrs) and their companions are generally non-degenerate and more
massive ($M_{c}\sim0.2-0.4M_{\odot}$). The most remarkable characteristic of RBs is
that these systems can possibly swinging between rotation-powered state and accretion-powered state. 
\footnote{The first case of a swinging pulsar PSR J1824-2452I was found in globular cluster M28 (Papitto et al. 2013).}
The prototypical example of this class in the Galactic field
is PSR~J1023+0038 (Archibald et al. 2009,2010), which was firstly identified as a low-mass X-ray binary (Homer et al. 2006) and subsequently
a radio MSP was discovered and a former accretion disk was found to disappear (Archibald et al. 2009). Interestingly, since 2013 late June,
the radio pulsation of this system has disappeared and a new disk has been formed, which indicates the system has re-entered the accretion-powered 
state (Stappers et al. 2014; Takata et al. 2014; Li et al. 2014). 

There are $\sim30\%$ of the known MSPs in the Galactic field are found to be isolated
(Manchester et al. 2005). As MSPs are considered to be the off-springs from the evolution of
compact binaries, the existence of isolated MSPs has raised a question for their origins. One possible explanation for their
solitudes is that their companions have been evaporated in the presence of the high energy radiation and/or the relativistic
wind particles from the companion MSPs (van den Heuvel \& van Paradijs 1988).

Statistical analyses of the X-ray properties of pulsars can place constraints on the radiation mechanisms 
(e.g. Seward \& Wang 1988; Becker \& Tr\"{u}mper 1997; Possenti et al. 2002; Kargaltsev \& Pavlov 2008; Vink et al. 2011; Prinz \& Becker 2015; Shibata et al. 2016).
Most of these studies focused on investigating the empirical relation between the X-ray luminosities $L_{x}$ and the spin-down power $\dot{E}$. 
The reported relations are found to diverse in the literature. This can be ascribed to a number of factors. First, the choice of energy bands can affect 
the best-fit relation. For those have included the soft band ($\lesssim2$~keV) in their study (e.g. Prinz \& Becker 2015), 
there can be contribution from the neutron star cooling in the young/middle-aged pulsars which should not covary with $\dot{E}$. Moreover, soft X-ray 
fluxes are more sensitive to the uncertainties due to the interstellar absorption. 

The diversity can also be related to the different sample selection criteria in different studies. Pulsar population is heterogeneous, which comprises 
many different subclasses: such as young pulsars (e.g. Crab), high magnetic-field pulsars, MSPs. The X-ray emission properties can be vary among different 
classes (e.g. Possenti et al. 2002; Shibata et al. 2016). Therefore, investigations that include all the X-ray detected pulsars in their samples are subjected to 
a large scattering. Therefore, recent statistical studies of pulsars are focused on particular sub-classes (e.g. Shibata et al. 2016). However, there 
is no corresponding updated analysis of MSPs. 

A detailed statistical analysis of X-ray detected MSPs dated back to Possenti et al. (2002) which 
suggest a best-fit relation of $\log L_{x}=(1.38\pm0.10)\log\dot{E}-(16.36\pm3.64)$
in 2-10~keV for ten MSPs with confirmed X-ray detections at that time (cf. Fig.~2 in Possenti et al. 2002).
%which is comparable with the relation deduced from the entire sample of X-ray detected pulsars adopted in their work, i.e.
%$\log L_{X}=1.34\log\dot{E}-15.34$ (cf. Fig.~2 in Possenti et al. 2002).

With a significantly enlarged sample, it is timely to re-examine the X-ray emission properties of MSPs. Also, as different sub-classes
of MSPs have now been identified, it is interesting to compare the physical and emission properties among different classes.
In this paper, we present the results from a detailed statistical analysis of the X-ray properties of MSPs in the Galactic field. 
For the MSPs residing in globular clusters, as the formation processes
are different from those in the Galactic field, their emission properties can possibly be different (Hui, Taam \& Cheng 2010).
Hence the globular cluster MSPs are excluded in this study.

\section{Data collection \& normalization}
Using ATNF pulsar catalog of the version in April 2017 (Manchester et al. 2005), we firstly compiled a list of radio pulsars based on the following criteria:

\begin{enumerate}
\item
Rotational period $P<20$~ms
\item
Excluding the pulsars in globular clusters
\end{enumerate}

There are 197 radio pulsars fulfilling the aforementioned criteria. We subsequently searched for the literature which is relevant
to the X-ray properties of the pulsars in our list. Since different studies have adopted different energy ranges and different spectral models in their
analyses, the published X-ray properties (such as $L_{x}$ and their spectral steepness) do not form a homogeneous set of data.

In order to construct a homogeneous data set for a meaningful statistical analysis, we normalized the data with the following procedures:
\begin{enumerate}
\item
We adopted a simple absorbed power-law (PL) model for all the X-ray detected MSPs. We emphasize that the photon indices $\Gamma$ adopted in
our analysis are not necessary reflecting the property of the non-thermal component as MSPs can also emit thermal X-rays from their
heated polar caps (Cheng \& Zhang 1999). However, for modeling the thermal component, while some studies have adopted a simple black-body model, 
some other have adopted more sophisticated model such as atmospheric model. In order to avoid this inhomogeneity, we used a simple PL with its $\Gamma$
as a effective index to provide a convenient measure of the X-ray hardness with the interstellar absorption corrected. 
\item
Using absorption-corrected X-ray fluxes $f_{x}$ and $\Gamma$ in the band reported in literature and their
corresponding statistical uncertainties, 
we computed $f_{x}$ and their errors for the X-ray detected MSPs in an energy range of $2-10$~keV with the aid of PIMMS.
\footnote{https://heasarc.gsfc.nasa.gov/docs/software/tools/pimms.html}
\item
For the distances $d$ used in calculating $L_{x}=4\pi d^{2}f_{x}$, more than half of the X-ray detected MSPs (25) have their estimates of $d$ derived
from the dispersion measures ($d_{DM}$) only 
(Manchester et al. 2005). For the MSPs which have their $d$ estimated by dedicated investigations (e.g. parallax),
these values are adopted instead of those derived from dispersion measures as they are more reliable.
We noted that most of our sample do not have any uncertainty estimate of $d$. To provide an overall uncertainty estimate of $d$,
we constructed the distribution for the relative difference between $d_{DM}$ and those determined by more accurate methods $(d-d_{DM})/d_{DM}$ of 
the entire pulsar population. The standard deviation of this distribution is found to be 0.41. This is consistent with the uncertainty of $\pm40\%$ as adopted by 
Possenti et al. (2002). In this study, we assume a percentage error of $\pm41\%$ for all our adopted distance which is subsequently propagated into 
the error budget of $L_{x}$. 

\item
We re-analysed the X-ray data of the MSPs and computed the $1\sigma$ uncertainties of $\Gamma$ and $L_{X}$ for the following cases: (i)
the X-ray properties of the MSPs are not modeled with a single PL in the existing literature (ii) there is no error estimation for either
$\Gamma$ or $L_{x}$ in the existing literature. For each of these MSPs, their background-subtracted X-ray spectra were fitted with an
absorbed PL model by using XSPEC.
For the MSPs with the spectral models different from a simple absorbed PL in literatures, we have compared their 
$f_{x}$ obtained in our analysis and their values reported in the corresponding literatures. 
And we found the differences are all less than $2\sigma$ of the flux uncertainties.
\end{enumerate}

In re-examining the {\it XMM-Newton} data of PSR~J1600-3053 and PSR~J1832-0836 (firstly reported by Prinz \& Becker 2015), 
we found the signal-to-noise ratios of the potential X-ray counterparts of these two MSPs are $\sim2\sigma$.
In view of this, these two pulsars will be excluded from our list of confirmed X-ray detected MSPs as both $\Gamma$ and $L_{x}$ cannot be
properly constrained.

In Table~1, we summarize the physical properties of 47  X-ray detected MSPs together with their $L_{x}$ and $\Gamma$ normalized by the aforementioned
procedures. Our MSP sample size is almost five times larger than that adopted in Possenti et al. (2002).

We have also found that 21 other MSPs without any reported X-ray investigation have been serendipitously covered by the archival data obtained by {\it Swift} XRT.
We have performed a systematic search for their X-ray counterparts by means of a wavelet source detection algorithm. For all these 21 MSPs, we do not
yield any detection larger than $3\sigma$. Estimating the $n_{H}$ of these MSPs by the total HI column density through the Galaxy at their radio
timing positions (Kalberla et al. 2005) and assuming a photon index of $\Gamma=2$,
we computed $1\sigma$ upper limits of their $L_{x}$ in 2-10~keV using their distances and limiting count rates.
Together with the limiting luminosities of 13 other MSPs reported by other literatures and those of PSR~J1600-3053 and PSR~J1832-0836
obtained from our re-analysis, the results are summarized in Table~2.

\section{Statistical Analysis}
With the sample prepared by the aforementioned procedures, we performed a detailed statistical analysis of the X-ray properties of MSPs. 

\subsection{Correlation \& regression analysis}
We started by investigating whether $L_{X}$ is correlated with a number of proper-motion corrected derived parameters, including spin-down power $\dot{E}$,
characteristic ages $\tau$, magnetic-field strength at the stellar surface $B_{s}$ and magnetic-field strength at the light cylinder
$B_{LC}$. 

We noted there is a large fraction of non-detections in our sample and these MSPs have relatively low $\dot{E}$ (see Table 2 and Figure 1), 
Ignoring these upper-limits will result in strong selection bias (i.e. Malmquist bias). 
For handling the censored data (i.e. 46 detections + 35 upper-limits), 
we have used the \texttt{CRAN NADA} package to perform the survival analysis\footnote{https://CRAN.R-project.org/package=NADA}.
Non-parametric approach has been adopted for both correlation and regression analysis. 

We used the generalized Kendall's $\tau$ for testing the correlation (cf. Helsel 2005). The corresponding $p-$values are summarized in Table~3.
Strong positive correlations are found for $L_{x}-\dot{E}$ ($p-$value=$2.4\times10^{-4}$) and $L_{x}-B_{LC}$ ($p-$value=$4.2\times10^{-4}$).
An anti-correlation between $L_{x}$ and $\tau$ ($p-$value=$4.6\times10^{-3}$) is also found. On the other hand, there is no evidence for the correlation
between $L_{x}$ and $B_{s}$ ($p-$value=0.97).

We proceeded with a linear regression analysis on the censored data to obtain an empirical $\log L_{x}-\log \dot{E}$ relation. The slope and the intercept are estimated 
by the Akritas-Thiel-Sen (ATS) line (Akritas et al. 1995) and the Turnbull estimate (Turnbull 1976) respectively. 
The $1\sigma$ errors of the model parameters are estimated by bootstrap resampling with 1500 simulated datasets. 

The best-fit relation is found to be
\begin{equation}
\log L_{x}=\left(1.31\pm0.22\right) \log\dot{E}-\left(14.80\pm7.68\right)
\end{equation}

Figure~1 clearly shows the positive correlation between $L_{x}$ of MSPs and their $\dot{E}$. The ATS fit is illustrated by a solid line. 
We have also carried out the standard linear regression with 
the X-ray detected MSPs only. This yields $\log L_{x}=1.26\log\dot{E}-12.53$. Together with the relation estimated by Possenti et al. (2002) with 10 MSPs, we overplot 
these two lines in Figure~1 for comparison. One should note that the ATS estimate lies below the other two lines. This demonstrates that the results of regression 
analysis without taking the upper-limits into account tend to overestimate the X-ray conversion efficiency. 

Applying the same procedures on the censored data, we have also obtained the best-fit relation for $L_{x}-B_{LC}$:
\begin{equation}
\log L_{x}=\left(1.63\pm0.30\right) \log B_{LC}+\left(22.12\pm1.45\right)
\end{equation}
\noindent and for $L_{x}-\tau$: 
\begin{equation}
\log L_{x}=-\left(1.53\pm0.42\right) \log\tau +\left(44.75\pm4.08\right).
\end{equation}

\subsection{Searches for differences among various MSP classes}
As MSPs can now be divided into different classes, it is interesting to compare their properties.
In our study, we divided our sample of X-ray detected MSPs into four classes: (1) RBs, (2) BWs, (3) Isolated MSP and
(4) Others. The classification of RBs and BWs is based on the online catalog provided by
Alessandro Patruno\footnote{https://apatruno.wordpress.com/about/millisecond-pulsar-catalogue/}.
For the MSPs which do not have any identified companions in the ATNF catalog (Manchester et al. 2005), we put them in the
category of isolated MSPs. For those not belonging to classes (1), (2) or (3), we put them into the forth category of ``Others".
The classes of each X-ray detected MSP are specified in Table~1.

To compare the properties among different MSP classes, we focus on the following parameters: $L_{x}$, $\Gamma$, $\tau$, $B_{LC}$, $B_{s}$, 
$\dot{E}$ and $P_{b}$. 

We firstly constructed the cumulative distribution functions of the aforementioned parameters, which are shown in Figures~2-8.
For searching the possible differences among these classes,
we apply a non-parametric two-sample Anderson-Darling (A-D) test (Anderson \& Darling 1952; Darling 1957; Pettitt 1976, Scholz \& Stephens 1987)
to their unbinned distributions.
We have also compared the results of A-D test with those obtained from the conventional Kolmogorov-Smirnov (K-S) test.
The results of both A-D test and K-S test are summarized in Table~4.

Among all the tested parameters, we found the distributions of $B_{s}$ of different classes are comparable.
On the other hand, the distributions of $B_{LC}$ and $\dot{E}$ for RBs are significantly different from those of ``Others"
($p-$value = 0.01 and 0.002 for $B_{LC}$ and $\dot{E}$ respectively) and marginally different from that of isolated MSPs 
($p-$value = 0.061 and 0.053 for $B_{LC}$ and $\dot{E}$ respectively).  Both $B_{LC}$ and $\dot{E}$ of RBs are found to be generally 
higher than those of isolated MSPs and ``Others".

In comparing the distributions of characteristic ages $\tau$, RBs are found to be younger than ``Others"
($p-$value$\sim0.0035$). On the other hand, there is a marginal indication that BWs are younger than ``Others"
($p-$value$\sim0.04$). 

The significances of the aforementioned results are
unaltered when the statistical uncertainties of the pulsar parameters given by the ATNF catalog are taken into account.

We have also compared the distributions of the orbital period $P_{b}$ for RBs, BWs and ``Others". The $P_{b}$ of Both RBs and BWs are 
significantly shorter than ``Others" ($p-$value$<5\times10^{-4}$), as expected given the selection criteria. On the other hand, there is no strong evidence that the 
$P_{b}$ distributions of RBs and BWs are different ($p-$value=0.1). 

While the distributions of the parameters (i.e. $\tau$, $B_{s}$, $B_{LC}$, $\dot{E}$ and $P_{b}$) of RBs are
comparable with those of BWs, their X-ray properties appears to be rather different. In comparing the
$L_{x}$ and $\Gamma$ distributions of these two classes, A-D tests yield the $p-$values of 0.003 and 0.01 respectively.
This suggests the X-ray emission of RBs is brighter and harder than that of BWs.

We would like to point out that the significances of their differences in
$L_{x}$ and $\Gamma$ can be reduced considerably ($p-$value=0.7 and 0.2 respectively) by taking into account the statistical uncertainties 
of the X-ray parameters. 

\section{Summary \& Discussion}
We have carried out a census for the X-ray population of MSPs and performed a detailed statistical analysis of their
physical and X-ray properties. $L_{x}$ is found to be positively correlated with $\dot{E}$ and $B_{LC}$. There is also evidence of
$L_{x}$ decreasing with $\tau$.
The best-fit $L_{x}-\dot{E}$ relation of MSPs are found to be $L_{x}\simeq 10^{31.05}\dot{E}_{35}^{1.31}$~erg/s in 2-10~keV,
where $\dot{E}$ is the spin-down power in units of $10^{35}$~erg/s. 

In comparing the properties among different classes of MSPs, we found that their distributions of $B_{s}$ are comparable.
However, $B_{LC}$ and $\dot{E}$ of RBs are significantly higher than those of ``Others". There is also some marginal
evidence that RBs and BWs are younger than ``Others".

While the pulsar and orbital parameters of RBs and BWs are comparable, $L_{x}$ of RBs are found to be higher than those
of BWs. Also, we have found the indication that the X-ray emission of RBs is harder than that of BWs in 2-10~keV though
a firm conclusion is limited by their statistical uncertainties.

In the following, we discuss the theoretical implication of these findings with a specific focus on RBs and BWs.

In MSP binary systems, there are three regions to produce the X-rays, which are intra-binary shock,  magnetosphere and heated polar cap. 
In the hypothesis that
  the GeV gamma-rays are produced in the outer magnetosphere around the light cylinder (Aliu et al. 2008; Abdo et al. 2013).
  the photon-photon pair-creation process makes a secondary pair. The synchrotron radiation of the
  secondary pairs can produce the synchrotron photon  with a typical photon energy of $E_{syn}\sim 0.7(\gamma_s/2\cdot 10^3)^2(B/10^5{\rm G})(\sin\theta/0.1)$keV, where $\gamma_s$ is the Lorentz factor of the secondary, $B$ is the magnetic field strength at the light cylinder and $\theta$ is the pitch angle (Takata et al. 2012).
  Based on the outer gap accelerator model,
  Takata et al. (2012) discuss that the magnetospheric synchrotron
  luminosity is related to the spin down power as $L_{syn}\sim 6\times 10^{30}(\dot{E}/10^{35}{\rm erg~s^{-1}})^{35/32}$, which is one or two orders of magnitude smaller than the observed X-ray emission from the BW/RB systems
  (green lines in Figures~\ref{bw1} and~\ref{rb1}).

  For the observed pulsed X-ray emission of the Fermi-LAT MSPs, the magnetospheric emission dominates in the spectral energy
  distribution of the MSPs with a higher spin down power, say $\dot{E}>10^{36}{\rm erg~s^{-1}}$
  (Ng et al. 2014).  For the MSPs with a lower spin down power,
  on the other hand,   the heated polar cap emission dominates the magnetospheric emission in the spectrum.  As we can see in
  Figures~\ref{bw1} and~\ref{rb1}, the binary systems with a smaller spin down power show a tendency to have very soft X-ray emission
  with an effective photon index $\Gamma>3$.  This index is difficult to be explained by a standard synchrotron emission from
  the shock with a typical power law index of the accelerated pulsar wind particles, and it suggests a contribution
  of the emission from the heated polar cap region.

  The heated polar cap emission from MSPs are observed with two components (Bogdanov \&  Grindlay 2009); (i) rim component
  with a temperature of $T_r\sim  5\times 10^5$K and
  an effective radius of $R_r\sim 3$km, and (ii) core component with a temperature of $T_c \sim 2\times 10^6$K and
  an effective  radius of $R_c\sim 0.1$km.   In
  Takata et al. (2012), the two temperatures are modeled as
  \begin{equation}
    T_{r}\sim 5.2\times 10^{5}\dot{E}_{35}^{7/96}B_{8}^{5/48}R^{-1/2}_{r,3}{\rm K},
\label{tr}
  \end{equation}
  and
  \begin{equation}
  T_{c}\sim 3.5\times 10^{6}\dot{E}_{35}^{3/32}B_{8}^{1/16}R^{-1/2}_{c,0.1}{\rm K},
\label{tc}
  \end{equation}
  respectively, where $\dot{E}_{35}=\dot E/10^{35}{\rm erg~s^{-1}}$, $R_{r,3}=R_r/3{\rm km}$ and $R_{c, 0.1}=R_c/0.1{\rm km}$.
  This heated polar cap emission will dominate  the shock emission in the observation of the BW with a lower spin down power
  (red lines in Figures~\ref{bw1}).

  To model the shock emission, we use a simple one-zone model and calculate the synchrotron radiation of the
  accelerated electrons/positrons in the pulsar wind. We assume that the shock is located close to the
  companion star, at a distance from the MSP $r_s\sim 10^{11}$cm is the typical separation between two stars.
  We evaluate the magnetic field of the pulsar wind at the shock from $B_s=3(\dot{E}\sigma_{PW}a/r_s^2c)^{1/2}\sim 17 \dot{E}^{1/2}_{35} (\sigma_{PW}/0.1)^{1/2}(r_s/10^{11}{\rm cm})^{-1}$G (Kennel \& Coroniti 1984), where $\sigma_{PW}$ is the ratio of the magnetic energy to the kinetic energy of the cold relativistic pulsar wind. 
    To compare with the population of BWs and RBw, we calculate the shock emission
    with $\sigma_{PW}=0.1$, at which the synchrotron emission from the shocked pulsar wind
    becomes maximum, and $r_s=10^{11}$cm.
  At the shock, the electrons/positrons are accelerated
  beyond the Lorentz factor ($\Gamma_1$) of the cold relativistic pulsar wind, and forms a power law distribution
  in the energy.  We assume that $\Gamma_1=10^5$ in the
  calculation. We determine the maximum Lorentz factor of the accelerated particles by balancing the
  accelerating time scale, $\tau_{acc}=\Gamma m_ec/(eB_s)$, and the synchrotron cooling time scale $\tau_s\sim 9m_ec^3c^5/(4e^4B_s^2\Gamma)$, that is, $\Gamma_{max}\sim 2.7\times 10^7(B_s/20{\rm G})^{-1/2}$. For the initial
  distribution, we assume the hard power low index $p=1.5$ to explain the hard spectrum of the X-ray emissions from
  some binary systems.   We solve the
  evolution of the distribution function under the synchrotron energy loss.By assuming the speed of the post shocked flow is constant, we calculate the magnetic field evolution from the magnetic flux conservation $B=B_s(r_s/r)$ (Kennel \& Coroniti 1984). 

  The shock luminosity is proportional to $L_{X}\propto \delta \dot{E}$, where $\delta$ is the fraction
  of the pulsar wind blocked by the outflow from the companion star and/or the companion star itself. 
  While the pulsar and orbital parameters of BWs and RBs are comparable,
  the X-ray luminosity of the RBs is significantly higher than that of the BWs.
This can be explained by the difference in the fraction $\delta$.
  We may estimate the fraction of the sky intercepted by the companion star with $\delta\sim (R_R/2a)^2$,
  where  $R_R$ is the Roche-lobe radius of the companion star. Since the
  Roche-lobe radius is estimated as $R_R/a=0.462[q/(1+q)]^{1/3}$ with $q$ being  the ratio of the
  companion mass to the
  neutron star mass (Frank et al. 2002), the fraction becomes $\delta=0.053[q/(1+q)]^{2/3}$. With the typical values of the mass ratio,
  we estimate   $\delta \sim 1$\% for RB ($q=0.1$) and $\delta \sim 0.2$\% for the BW ($q=0.01$). For most RBs, moreover,
  the companion star has been identified as a low-mass (G/M type) main-sequence star (see ATNF catalog, Manchester et al. 2005). In such system, 
  a larger fraction could be realized  by the magnetized outflow from the companion star (Archibald et al. 2013).
  Since the companion star is tidally locked, the spin period is equal to the orbital period of several hours. The low mass, rapidly
  spinning main sequence star can have a surface magnetic field of several hundreds to a few thousand Gauss
  (Reiners et al. 2009).
   Since the shock distance will be located at several stellar radius
    of the companion star, the magnetized outflow from the companion star
    may overcome the pulsar
    wind. We may  assume the radial dependence of the stellar magnetic field as $B(R)=B_*(R_*/R)^{m}$, where $B_*$ and $R_*$ are stellar magnetic field and
    radius, respectively, and $R$ is the distance from the center of the star.
    The magnetic field  in  the stellar wind could deviate
    from the dipole field ($m=3$) and have $m<3$, owing to the spin of
    the star and/or more complicate surface magnetic field structure
    (Banaszkiewicz et al. 1998). The typical distance to the shock from
    the pulsar, $r_s$, may be estimated from the momentum balance of
    $B^2(R)c=L_{sd}/r_s$, where $r_s=a-R$. Then we would estimate required surface
    magnetic field strength
    to overcome the pulsar wind from the condition that $r_s<a/2$. Under
    aforementioned magnetic field configuration, we have $B_*>(a/2)^{m-1}R_*^{-m}(L_{sd}/c)^{1/2}$, which yields $B_*\gtrsim 500{\rm G}$ for the dipole field ($m=3$) and $B_*\gtrsim 200$G for $m=2$ with $a=10^{11}{\rm cm}$, $L_{sd}=10^{35}{\rm erg/s}$ and $R_*=0.3R_{\odot}$. Hence, the magnetized stellar wind will be able to
    overcome the pulsar wind if the stellar magnetic field  can be enhanced to $B_*>10^{2-3}$G.
  %We may assume that the radial dependency of the magnetic field is described by
%    $B(R)=B_*(R_*/R)^2$, where $B_*$ and $R_*$ are the surface
%    magnetic field and radius of the companion star, and $R$ is the distance f%rom the center of the star to the shock. Then we
%    can estimate the ration of the momenta for the magnetized stellar wind to %the pulsar wind as
%    $\eta=B^2_*R_*c/L_sd\sim 1(B_*/10^2{\rm G})^2[R_*/(0.3R_{\odot})]^2(L_{sd}%/10^{35}{\rm erg~s^{-1}})^{-1}$. Hence,
%    the magnetized stellar wind will overcome the pulsar wind if the stellar magnetic field can be enhanced to $B_*>10^2$G. }

  It has been observed that the low frequency radio wave for the RB  is observed with an eclipse lasting a large part of
  the orbital phase. This also suggests that a large fraction of the pulsar's sky is blocked by the outflow from the companion stars.

  The predicted luminosity and photon index from BW and RB are summarized in
  the Figures~\ref{bw1}$\sim$\ref{rb2}; In each figure, the blue lines show the model
  prediction for the relation between  the X-ray luminosity and  spin down power
  (left  panels) or the photon index (right panels), and they consider
  the contribution from  the shock emission and the heated polar cap emission. In Figure~\ref{bw1} and~\ref{rb1}, we consider the different fraction, $\delta$, for the shock emission; the solid, dashed and dotted lines are results for $\delta=$0.15\%, 0.3\% and 0.6\%, respectively. We can see in the figure that for a fixed X-ray luminosity, the calculated photon index is softer for a smaller efficiency, $\delta$. This is because the contribution
  of the shock emission relatively to the heated polar cap emission becomes smaller. For the BW, our model predicts that the X-ray emission  for
  the spin down power  $\dot{E}<10^{34}{\rm erg~s^{-1}}$ is dominated by the heated polar cap emission. This can explain the observed soft spectrum of some BW pulsars. As  Figure~\ref{rb1} shows,
  the current model predicts that the observed X-ray emissions from the RB is dominated by
  the shock emission, and the contribution of the heated polar cap emission is negligible, except for
   PSR~J1816+4516, for which the heated polar cap emission may be observed.  
   In Figures~\ref{bw2} and ~\ref{rb2}, we summarize the dependency on the surface magnetic
   field, which affects the heated polar
   cap temperature through the equations (\ref{tr}) and~(\ref{tc}); the solid, dashed and dotted
   lines are results for $B_s=1.5\times 10^8$G, $3\times 10^8$G and $6\times 10^8$G, respectively. 
   We can see in the figures that for a  fixed spin down power, the predicted X-ray emission
   shows a  less dependency on the surface magnetic field. With a reasonable range of the parameters
    of the MSPs, the current model is qualitatively consistent
   with the observations. Based on our result, we suggest  the pulsation search for the BWs with a lower
   spin down power, for which the emission will be dominated by the heated polar cap emission, since  an increase in the sample of
   the pulse profiles of the heated polar cap will be useful to  study the equation of the state for the neutron star.

\acknowledgements
JL is supported by BK21 plus Chungnam National University and the National Research Foundation of Korea grant 2016R1A5A1013277;
CYH is supported by the National Research Foundation of Korea grant 2016R1A5A1013277;
JT is supported by the NSFC grants of China under 11573010, U1631103 and 11661161010;
AKHK by the Ministry of Science and Technology of Taiwan
    grants 106-2918-I-007-005 and 105-2112-M-007-033-MY2;
PHT is supported by the National Natural Science Foundation of China (NSFC) grants 11633007 and 11661161010;
KSC are supported by GRF grant under 17302315.

\clearpage
\begin{table}
\begin{center}
\scriptsize
\caption{Pulsar parameters and X-ray properties of 47 X-ray detected MSPs. $\dot{E}$, $\tau$, $B_{\rm s}$ and $B_{\rm LC}$ are derived from the
proper-motion corrected period derivatives.}
\begin{tabular}{c c c c c c c c c c c c c}
\hline
\hline
MSP name & Inst. & Class & $P$ & $P_{b}$ & $\log\dot{E}$ & $d$ & $\tau$ & $B_{\rm s}$ & $B_{\rm LC}$ & $\log L_{x}$ (2-10 keV) & $\Gamma$  & References \\
 & & & ms & day & erg/s& kpc & Gyr & $10^8$ G & $10^4$ G & erg/s & & \\
\hline
J0023+0923 & C & BW & 3.05 & 0.14 & 34.13 & $1.25\pm0.52$ & 5.03 & 1.73 & 5.55 & 29.35$^{+0.47}_{-0.53}$ & $3.3\pm0.5$ & 1 \\
J0030+0451 & R & I & 4.87 & $\cdot\cdot$ & 33.53 & $0.36\pm0.15$ & 7.69 & 2.24 & 1.76 & 29.88$^{+0.84}_{-1.19}$ & $2.0\pm0.2$ & 2 \\
J0034-0534 & X & O & 1.88 & 1.589 & 34.38 & $1.35\pm0.56$ & 7.32 & 0.88 & 12.10 & 29.48$^{+0.78}_{-1.11}$ & $2.75^{+1.29}_{-0.70}$ & 3 \\
J0101-6422 & C & O & 2.57 & 1.788 & 33.93 & $1.00\pm0.41$ & 11.2 & 0.98 & 5.23 & 29.90$^{+0.38}_{-0.54}$ & $3.25^{+0.56}_{-0.55}$ & 3,4 \\
J0218+4232 & B & O & 2.32 & 2.029 & 35.38 & $3.15\pm1.30$ & 0.48 & 4.27 & 31.06 & 33.20$^{+0.62}_{-0.93}$ & $0.94\pm0.22$ & 2 \\
J0337+1715 & X & O & 2.73 & 1.629 & 34.53 & $1.30\pm0.54$ & 2.45 & 2.22 &10.2 & 29.76$^{+0.47}_{-0.64}$ & $3.6^{+1.1}_{-0.8}$ & 5 \\
J0437-4715 & R & O & 5.76 & 5.741 & 33.45 & $0.16\pm0.07$ & 6.64 & 2.85 & 1.35 & 30.19$^{+0.51}_{-0.71}$ & $2.35\pm0.35$ & 2 \\
J0613-0200 & X & O & 3.06 & 1.199 & 34.09 & $0.78\pm0.32$ & 5.42 & 1.68 & 5.31 & 29.50$^{+0.67}_{-1.01}$ & $2.7\pm0.4$ & 6 \\
J0614-3329 & S & O & 3.15 & 53.59 & 34.34 & $0.62\pm0.26$ & 2.85 & 2.38  & 6.91 & 29.92$^{+0.35}_{-0.52}$ & $2.63^{+0.30}_{-0.27}$ & 7 \\
J0636+5129 & X & O & 2.87 & 0.07 & 33.75 & $0.20\pm0.08$ & 13.6 & 0.99 & 3.82 & 27.93$^{+1.18}_{-0.82}$ & $5.0^{+5.0}_{-1.0}$ & 5 \\
%//
J0751+1807 & R & O & 3.48 & 0.26 & 33.75 & $1.11\pm0.46$ & 9.14 & 1.47 & 3.16 & 31.29$^{+0.85}_{-1.08}$ & $2.0\pm0.2$ & 2 \\
J1012+5307 & R & O & 5.26 & 0.61 & 33.49 & $0.70\pm0.29$ & 7.39 & 2.6 & 1.54 & 29.58$^{+0.83}_{-1.19}$ & $2.3\pm0.2$ & 2 \\
J1023+0038 & C & RB & 1.96 & 0.20 & 34.4. & $1.37\pm0.57$ & 2.63 & 0.94 & 12.24 & 31.90$^{+0.31}_{-0.75}$ & $1.19\pm0.03$ & 8 \\
J1024-0719 & R & O & 5.16 & $\cdot\cdot$ & 33.72 & $1.22\pm0.50$ & 4.41 & 3.13 & 2.13 & 29.09$^{+0.80}_{-1.25}$ & $2.0\pm0.2$ & 2 \\
J1124-3653$^*$ & C & BW & 2.41 & $\cdot\cdot$ & 33.6 & $1.05\pm0.43$ & 27.05 & 0.59 & 3.88 & 30.56$^{+0.45}_{-0.68}$ & $2.1\pm0.3$ & 9 \\
J1227-4853 & C,X & RB & 1.69 & 0.29 & 34.96 & $1.80\pm0.74$ & 2.41 & 1.38 & 27.1 & 32.12$^{+0.31}_{-0.47}$ & $1.2\pm0.04$ & 10 \\
J1300+1240 & C & O & 6.22 & 25.26 & 33.70 & $0.60\pm0.25$ & 3.23 & 4.41 & 1.67 & 28.82$^{+0.38}_{-0.56}$ & $2.75\pm0.35$ & 11 \\
J1311-3430 & C & BW & 2.56 & 0.07 & 34.69 & $2.43\pm1.00$ & 1.94 & 2.34 & 13.1 & 31.63$^{+0.37}_{-0.54}$ & $1.3\pm0.3$ & 1 \\
J1417-4402 & C & O & 2.66 & 5.37 & $\cdot\cdot$ & $4.40\pm1.82$ & $\cdot\cdot$ & $\cdot\cdot$ & $\cdot\cdot$ & 33.14$^{+0.38}_{-0.61}$ & $1.32\pm0.4$ & 12 \\
J1446-4701 & X & BW & 2.19  & 0.28 & 34.56 & $1.57\pm0.65$ & 3.61 & 1.47 & 12.73 & 30.30$^{+0.72}_{-0.75}$ & $2.9^{+0.5}_{-0.4}$ & 1 \\
%//
J1514-4946 & C & O & 3.59 & 1.92 & 33.99 & $0.91\pm0.38$ & 4.94 & 2.06 & 4.04 & 29.37$^{+0.46}_{-0.61}$ & $2.98^{+1.2}_{-0.99}$ & 3 \\
J1614-2230 & X & O & 3.15 & 8.69 & 33.70 & $0.70\pm0.29$ & 12.5 & 1.13 & 3.30 & 27.55$^{+0.35}_{-0.56}$ & $4.64^{+1.50}_{-0.88}$ & 13 \\
J1628-3205$^*$ & C & RB & 3.21 & 0.21 & 34.26$^+$ & $1.22\pm0.50$ & 3.38 & 2.23 & 6.20 & 30.96$^{+0.33}_{-0.50}$ & $1.88^{+0.20}_{-0.19}$ & 3 \\
J1640+2224 & C & O & 3.16 & 175.46 & 33.20 & $1.50\pm0.62$ & 39.0 & 0.64 & 1.85 & 29.64$^{+0.38}_{-0.55}$ & $3.02^{+0.50}_{-0.46}$ & 3 \\
J1658-5324 & C & I & 2.44 & $\cdot\cdot$ & 34.48 & $0.88\pm0.36$ & 3.50 & 1.66 & 10.38 & 29.64$^{+0.39}_{-0.56}$ & $3.22^{+0.65}_{-0.64}$ & 3,4 \\
J1709+2313 & C & O & 4.63 & 22.71 & 32.63 & $2.18\pm0.90$ & 68.5 & 0.713 & 0.65 & 30.05$^{+0.44}_{-0.62}$ & $1.49^{+0.86}_{-0.80}$ & 3 \\
J1723-2837 & C,X & RB & 1.86 & 0.62 & 34.67 & $0.72\pm0.30$ & 3.9 & 1.2 & 17.6 & 31.92$^{+0.31}_{-0.48}$ & $1.0\pm0.07$ & 14 \\
J1730-2304 & X & I & 8.12 & $\cdot\cdot$ & 33.02 & $0.62\pm0.26$ & 8.99 & 3.45 & 0.59 & 29.18$^{+0.48}_{-0.63}$ & $2.7^{+0.9}_{-0.5}$ & 4 \\
J1731-1847 & C & BW & 2.34 & 0.31 & 34.87 & $4.78\pm1.98$ & 1.53 & 2.42 & 17.12 & 30.99$^{+0.60}_{-0.80}$ & $1.9^{+1.5}_{-1.3}$ & 1 \\
J1744-1134 & C & I & 4.07 & $\cdot\cdot$ & 33.63 & $0.40\pm0.17$ & 8.96 & 2.34 & 2.68 & 29.09$^{+0.51}_{-0.718}$ & $2.0\pm0.2$ & 2 \\
%//
J1810+1744$^*$ & C & BW & 1.66 & 0.15 & 34.60 & $2.36\pm0.98$ & 5.72 & 0.88 & 17.80 & 30.68$^{+0.44}_{-0.52}$ & $2.2\pm0.4$ & 1 \\
J1816+4510 & C & RB & 3.19 & 0.36 & 34.71 & $4.36\pm1.80$ & 1.21 & 3.70 & 10.35 & 30.32$^{+0.51}_{-0.59}$ & $2.76^{+0.74}_{-0.69}$ & 3,4 \\
J1909-3744 & C & O & 2.95 & 1.53 & 34.64 & $1.14\pm0.47$ & 16.5 & 0.92 & 3.27 & 29.72$^{+0.36}_{-0.51}$ & $3.02^{+0.40}_{-0.39}$ & 3 \\
J1911-1114 & X & O & 3.63 & 2.72 & 33.97 & $1.07\pm0.44$ & 5.04 & 2.06 & 3.91 & 30.40$^{+0.34}_{-0.51}$ & $1.89^{+0.20}_{-0.19}$ & 3 \\
J1939+2134 & B & I & 1.56 & $\cdot\cdot$ & 36.04 & $3.50\pm1.45$ & 0.24 & 4.09 & 98.08 & 32.78$^{+0.36}_{-0.53}$ & $1.94^{+0.13}_{-0.11}$ & 15 \\
J1959+2048 & C & BW & 1.61 & 0.38 & 35.00 & $1.73\pm0.72$ & 2.40 & 1.32 & 28.80 & 31.04$^{+0.32}_{-0.48}$ & $1.96\pm0.12$ & 16 \\
J2017+0603 & C & O & 2.9 & 2.20 & 34.11 & $1.40\pm0.58$ & 5.77 & 1.54 & 5.73 & 29.98$^{+0.40}_{-0.57}$ & $2.75\pm0.71$ & 3,4 \\
J2043+1711 & C & O & 2.38 & 1.48 & 34.08 & $1.25\pm0.52$ & 9.11 & 1.00 & 6.77 & 29.79$^{+0.46}_{-0.64}$ & $3.03^{+2.14}_{-2.28}$ & 3 \\
J2047+1053$^*$ & C & BW & 4.29 & 0.12 & 34.02 & $2.79\pm1.15$ & 3.24 & 3.04 & 3.54 & 30.99$^{+0.52}_{-0.70}$ & $0.87\pm0.68$ & 1 \\
J2051-0827 & C,X & BW & 4.51 & 0.099 & 33.72 & $1.47\pm0.61$ & 5.89 & 2.37 & 2.34 & 28.62$^{+0.55}_{-0.70}$ & $4.1\pm0.7$ & 1 \\
%//
J2124-3358 & R & I & 4.93 & $\cdot\cdot$ & 33.38 & $0.41\pm0.17$ & 10.7 & 1.92 & 1.45 & 29.77$^{+0.62}_{-1.14}$ & $2.0\pm0.2$ & 2 \\
J2129-0429$^*$ & X & RB & 7.62 & 0.64 & 34.48 & $1.83\pm0.76$ & 0.36 & 16.20 & 3.37 & 31.79$^{+0.34}_{-0.51}$ & $1.25\pm0.04$ & 17 \\
J2214+3000 & C & BW & 3.12 & 0.42 & 34.22 & $0.60\pm0.25$ & 3.88 & 2.02 & 6.03 & 28.78$^{+0.48}_{-0.62}$ & $3.8\pm0.4$ & 1 \\
J2215+5135 & C & RB & 2.61 & 0.17 & 34.87 & $2.77\pm1.15$ & 1.24 & 2.99 & 15.80 & 31.92$^{+0.41}_{-0.61}$ & $1.4\pm0.2$ & 9 \\
J2241-5236 & C & BW & 2.19 & 0.15 & 34.40 & $0.96\pm0.40$ & 5.22 & 1.22 & 10.90 & 29.88$^{+0.42}_{-0.57}$ & $2.8\pm0.4$ & 1 \\
J2256-1024$^*$ & C & BW & 2.29 & 0.21 & 34.60 & $1.33\pm0.55$ & 3.00 & 1.69 & 12.92 & 30.08$^{+0.37}_{-0.53}$ & $2.9\pm0.3$ & 1 \\
J2339-0533 & S & RB & 2.88 & 0.19 & 34.04 & $1.10\pm0.45$ & 6.83 & 1.41 & 5.35 & 31.44$^{+0.33}_{-0.49}$ & $1.32\pm0.08$ & 18 \\
\hline
\hline
\end{tabular}
\end{center}
\small
\vspace{-0.5cm}
\tablecomments {\footnotesize Inst: C, X, S, Sw, R, and B stand for {\it Chandra, XMM-Newton, Suzaku, Swift, ROSAT, and BeppoSAX}, respectively. |
Class: I, BW, RB and O stand for isolated MSPs, black-widows, redbacks and others respectively. |
References : (1)Arumugasamy et al. (2015), (2)Possenti et al. (2002), (3)this work, (4)Prinz et al. (2015), (5)Spiewak et al. (2016), (6)Marelli et al. (2011), (7)Aoki et al. (2012), (8)Bogdanov et al. (2011), (9)Gentile et al. (2014), (10)Bogdanov et al. (2014), (11)Pavlov et al. (2007), (12)Strader et al. (2015), (13)Pancrazi et al. (2012), (14)Hui et al. (2014), (15)Nicastro et al. (2004), (16)Huang et al. (2012), (17)Hui et al. (2015), (18)Yatsu et al. (2015). |
$\dagger$ $\dot{E}$ is obtained from Roberts et al. (2015). *$\dot{P}$, $B_S$, $B_{LC}$, and $\tau$ are deduced from $\dot{E}$ obtained in the corresponding 
references. Pulsar Parameters of all the others are obtained from ATNF catalog (Manchester et al. 2005).
}
\end{table}

\clearpage
\begin{table}
\begin{center}
\scriptsize
\caption{Upper limits of $L_{x}$ of 36 MSPs. $\dot{E}$, $\tau$, $B_{\rm s}$ and $B_{\rm LC}$ are derived from the 
proper-motion corrected period derivatives.}
\begin{tabular}{c c c c c c c c c c c c}
\hline
\hline
MSP name & Inst. & Class & $P$ & $P_{b}$ & $\log\dot{E}$ & $d$ & $\tau$ & $B_{\rm s}$ & $B_{\rm LC}$ & $\log L_{x}$ (2-10 keV) & References \\
 & & & ms & day & erg/s & kpc & Gyr & $10^8$ G & $10^4$ G & erg/s & \\
\hline
J0340+4130 & Sw & I & 3.30 & $\cdot\cdot$ & 33.87 & $1.60\pm0.66$ & 7.76 & 1.51 & 3.82 & $<$30.99 & 1 \\
J0610-2100 & Sw & BW & 3.86 & 0.286 & 33.92 & $3.26\pm1.34$ & 51.20 & 0.69 & 1.09 & $<$31.70 & 2 \\
J0645+5158 & X & I & 8.85 & $\cdot\cdot$ & 32.35 & $0.80\pm0.33$ & 35.80 & 1.89 & 0.25 & $<$29.74 & 3 \\
J1022+1001 & Sw & O & 16.45 & 7.81 & 32.45 & $1.13\pm0.46$ & 8.16 & 7.34 & 0.15 & $<$30.270 & 1 \\
J1048+2339 & Sw & RB & 4.67 & 0.25 & 33.89 & $2.00\pm0.82$ & 3.68 & 3.10 & 2.77 & $<$31.17 & 1 \\
J1103-5403 & C & I & 3.39 & $\cdot\cdot$ & 33.57 & $1.68\pm0.69$ & 14.60 & 1.13 & 2.72 & $<$30.41 & 3 \\
J1435-6100 & Sw & O & 9.35 & 1.36 & 33.08 & $2.81\pm1.15$ & 6.05 & 4.84 & 0.56 & $<$31.88 & 1 \\
J1455-3330 & Sw & O & 7.99 & 76.18 & 33.25 & $1.01\pm0.41$ & 5.50 & 4.34 & 0.77 & $<$30.65 & 1 \\
J1525-5545 & Sw & O & 11.36 & 0.99 & 33.54 & $3.14\pm1.29$ & 1.37 & 12.40 & 0.79 & $<$31.11 & 1 \\
J1544+4937 & Sw & BW & 2.16 & 0.12 & 34.08 & $2.99\pm1.23$ & 11.70 & 0.81 & 7.50 & $<$31.28 & 1 \\
J1600-3053 & X & O & 3.6 & 14.35 & 33.87 & $1.80\pm0.74$ & 6.53 & 1.79 & 3.50 & $<$30.70 & 1 \\
J1643-1224 & X & O & 4.62 & 147.02 & 33.86 & $0.74\pm0.30$ & 4.06 & 2.92 & 2.69 & $<$30.75 & 3 \\
J1713+0747 & Sw & O & 4.57 & 67.83 & 33.52 & $1.18\pm0.48$ & 9.04 & 1.94 & 1.84 & $<$31.05 & 2 \\
J1719-1438 & X & O & 5.79 & 0.09 & 33.18 & $0.34\pm0.14$ & 12.30 & 2.10 & 0.98 & $<$29.37 & 3 \\
J1738+0333 & Sw & O & 5.85 & 0.36 & 33.65 & $1.47\pm0.60$ & 4.11 & 3.68 & 1.67 & $<$30.54 & 1 \\
J1741+1351 & Sw & O & 3.75 & 16.34 & 34.34 & $1.08\pm0.44$ & 2.05 & 3.33 & 5.75 & $<$30.38 & 1 \\
J1745-0952 & Sw & O & 19.38 & 4.94 & 32.67 & $0.23\pm0.09$ & 3.56 & 13.10 & 0.16 & $<$29.16 & 1 \\
J1745+1017 & Sw & BW & 2.65 & 0.73 & 33.68 & $1.21\pm0.50$ & 18.60 & 0.78 & 3.81 & $<$30.57 & 1 \\
J1748-3009 & C & O & 9.68 & 2.93 & $\cdot\cdot$ & $5.07\pm2.08$ & $\cdot\cdot$ & $\cdot\cdot$ & $\cdot\cdot$ & $<$32.83 & 3 \\
J1751-2857 & C,X & O & 3.91 & 110.75 & 33.84 & $1.09\pm0.45$ & 5.94 & 2.05 & 3.11 & $<$30.81 & 3 \\
J1804-2717 & Sw & O & 9.34 & 11.13 & 33.24 & $0.80\pm0.33$ & 4.17 & 5.83 & 0.65 & $<$30.09 & 1 \\
J1811-2405 & Sw & O & 2.66 & 6.27 & 34.42 & $1.83\pm0.75$ & 3.15 & 1.84 & 8.88 & $<$31.40 & 1 \\
J1832-0836 & X & I & 2.72 & $\cdot\cdot$ & 34.23 & $0.81\pm0.33$ & 5.0 & 1.55 & 7.22 & $<$30.40 & 1 \\
J1843-1113 & X & I & 1.85 & $\cdot\cdot$ & 34.77 & $1.26\pm0.52$ & 3.09 & 1.34 & 19.36 & $<$30.62 & 3 \\
J1850+0124 & Sw & O & 3.56 & 84.95 & 33.98 & $3.39\pm1.39$ & 5.18 & 1.99 & 4.14 & $<$32.35 & 1 \\
J1853+1303 & X & O & 4.09 & 115.65 & 33.69 & $1.32\pm0.54$ & 7.57 & 1.89 & 2.51 & $<$30.64 & 3 \\
J1857+0943 & X & O & 5.36 & 12.33 & 33.65 & $1.20\pm0.49$ & 4.92 & 3.08 & 1.82 & $<$30.61 & 3 \\
J1900+0308 & Sw & O & 4.91 & 12.48 & 33.30 & $4.80\pm1.97$ & 13.20 & 1.72 & 1.37 & $<$32.49 & 1 \\
J1901+0300 & Sw & O & 7.80 & 2.40 & 33.58 & $5.29\pm2.17$ & 2.70 & 6.04 & 1.19 & $<$31.94 & 1 \\
J1903-7051 & Sw & O & 3.60 & 11.05 & 33.82 & $0.93\pm0.38$ & 5.46 & 1.69 & 3.29 & $<$30.45 & 1 \\
J1933-6211 & C & O & 3.54 & 12.82 & 33.43 & $0.65\pm0.27$ & 15.20 & 1.05 & 2.15 & $<$30.28 & 3 \\
J1943+2210 & Sw & O & 5.08 & 8.31 & 33.42 & $6.78\pm2.78$ & 9.17 & 2.14 & 1.53 & $<$32.54 & 1 \\
J1946+3417 & X & O & 3.17 & 27.02 & 33.59 & $6.97\pm2.86$ & 16.10 & 1.01 & 2.96 & $<$32.38 & 3 \\
J2145-0750 & Sw & O & 16.05 & 6.84 & 32.40 & $0.53\pm0.22$ & 9.69 & 6.57 & 0.14 & $<$29.72 & 1 \\
J2229+2643 & Sw & O & 2.98 & 93.02 & 33.19 & $1.80\pm0.74$ & 45.30 & 0.56 & 1.94 & $<$30.95 & 1 \\
J2317+1439 & Sw & O & 3.45 & 2.46 & 33.34 & $1.43\pm0.59$ & 24.20 & 0.89 & 1.99 & $<$30.75 & 1 \\
\hline
\hline
\end{tabular}
\end{center}
\small
\tablecomments {Inst: C, X, and Sw stand for {\it Chandra, XMM-Newton, and Swift} respectively. |
Class: I, BW, RB and O stand for isolated MSPs, black-widows, redbacks and others respectively.
| References:
(1) this work, (2) Espinoza et al. 2013, (3) Prinz et al. 2015}
\end{table}

\begin{table}
\begin{center}
\caption{Summary of non-parametric correlation analysis between $L_{x}$ and various pulsar parameters with the censored data.}
\begin{tabular}{c c c}
\hline
\hline
 & Kendall's $\tau$ & $p$-value \\
 \hline
$\dot{E}$ & 0.27 & $2.4\times10^{-4}$ \\
$B_{LC}$ & 0.26 & $4.2\times10^{-4}$ \\
$B_S$ & $-3.1\times10^{-3}$ & 0.97 \\
$\tau$ & -0.21 & $4.6\times10^{-3}$ \\
\hline
\hline
\end{tabular}
\end{center}
%\small
%\tablecomments {The results of Spearman rank test.}
\end{table}

\begin{table}
\begin{center}
\scriptsize
\caption{Summary of the significances ($p$-value) of K-S and A-D tests.}
\begin{tabular}{c | cc | cc | cc | cc | cc | cc | cc}
\hline
\hline
 & \multicolumn{2}{c|}{$B_{\rm S}$} & \multicolumn{2}{c|}{$B_{\rm LC}$} & \multicolumn{2}{c|}{$\dot{E}$} & \multicolumn{2}{c|}{$\tau$} & \multicolumn{2}{c|}{$L_x$} & \multicolumn{2}{c}{$\Gamma$} & \multicolumn{2}{c}{$P_{b}$}\\
 & KS & AD & KS & AD & KS & AD & KS & AD & KS & AD & KS & AD & KS & AD \\
\hline
I vs. O & 0.14 & 0.22 & 0.35 & 0.28 & 0.19 & 0.26 & 0.72 & 0.51 & 0.85 & 0.96 & 0.41 & 0.35 & $\cdot\cdot$ & $\cdot\cdot$ \\
I vs. BWs & 0.38 & 0.18 & 0.080 & 0.028 & 0.080 & 0.046 & 0.080 & 0.15 & 0.19 & 0.27 & 0.67 & 0.49 & $\cdot\cdot$ & $\cdot\cdot$ \\
I vs. RBs & 0.25 & 0.40 & 0.051 & 0.061 & 0.051 & 0.053 & 0.051 & 0.033 & 0.0065 & 0.012 & 0.0036 & 0.0077 & $\cdot\cdot$ & $\cdot\cdot$ \\
O vs. BWs & 0.75 & 0.72 & 0.046 & 0.015 & 0.013 & 0.007 & 0.052 & 0.044 & 0.087 & 0.14 & 0.83 & 0.86 & 0.000003 & 0.00008 \\
O vs. RBs & 0.71 & 0.52 & 0.036 & 0.011 & 0.005 & 0.002 & 0.0052 & 0.0035 & 0.0003 & 0.0009 & 0.0017 & 0.0020 & 0.0001 & 0.0005 \\
BWs vs. RBs & 0.73 & 0.53 & 1.00 & 1.00 & 0.73 & 0.52 & 0.41 & 0.17 & 0.014 & 0.0032 & 0.0074 & 0.010 & 0.081 & 0.10 \\
\hline
\hline
\end{tabular}
\end{center}
\small
\tablecomments {I, BW, RB and O stand for isolated MSPs, black-widows, redbacks and Others respectively.}
\end{table}

%\clearpage
\begin{figure}
\centering
\includegraphics[width=
6.5in, angle=0]{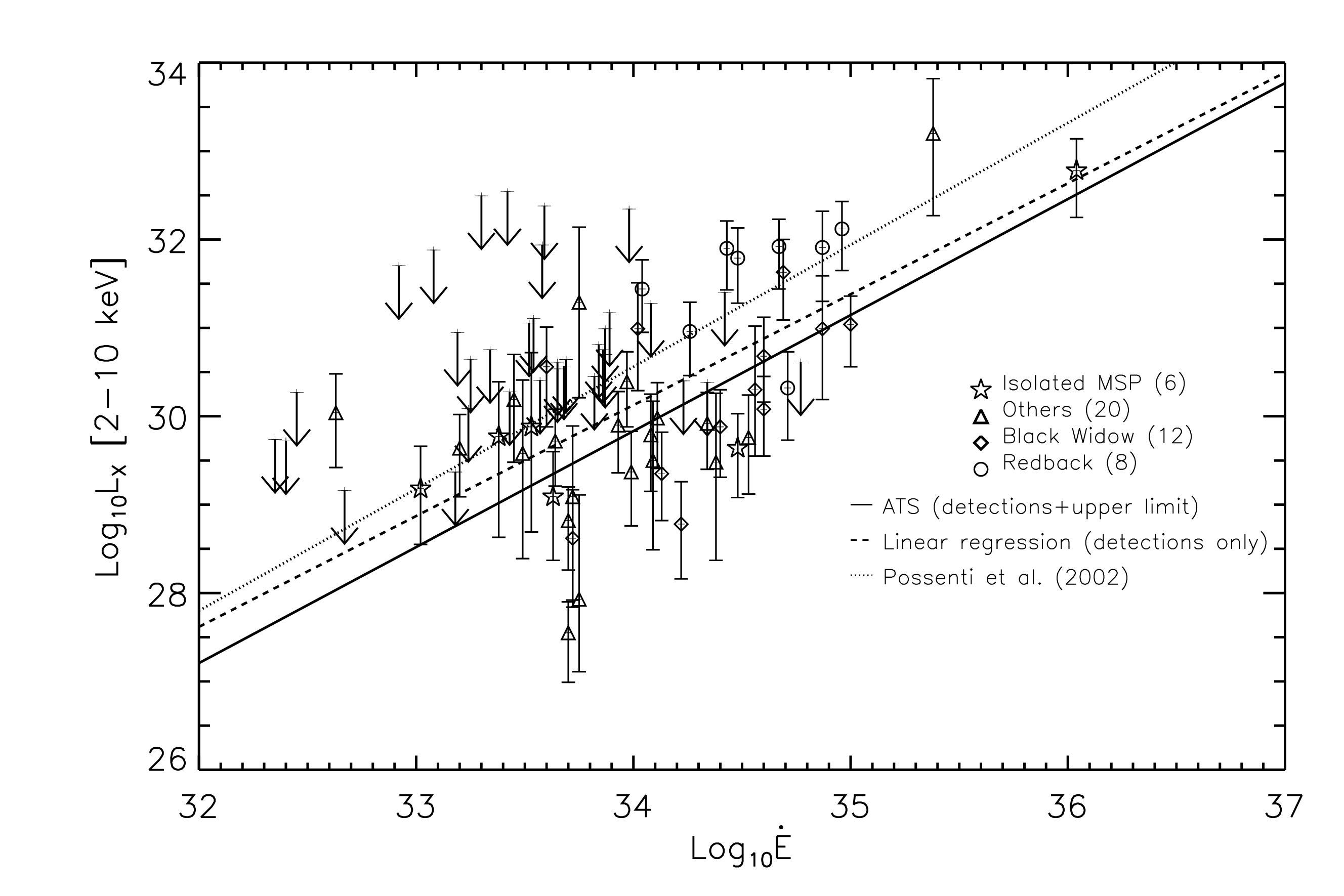}
\caption{Plot of $L_{x}$ vs $\dot{E}$ for MSPs. Different symbols represent different classes of 46 X-ray detected MSPs with measured $\dot{E}$ (cf. Tab.~1). 
The arrows illustrate the upper-limits on $L_{x}$ for 35 MSPs with measured $\dot{E}$ given in Tab.~2. The solid line shows the ATS line (Equation 1) as inferred from 
the censored data. For comparison, we also plotted the relation obtained from the standard linear regression of X-ray detected MSPs only (dashed line) 
and that reported by Possenti et al. (2002) based on a sample of 10 MSPs (dotted line).} 
\end{figure}

\clearpage
\begin{figure}
\centering
\includegraphics[width=
6.0in, angle=0]{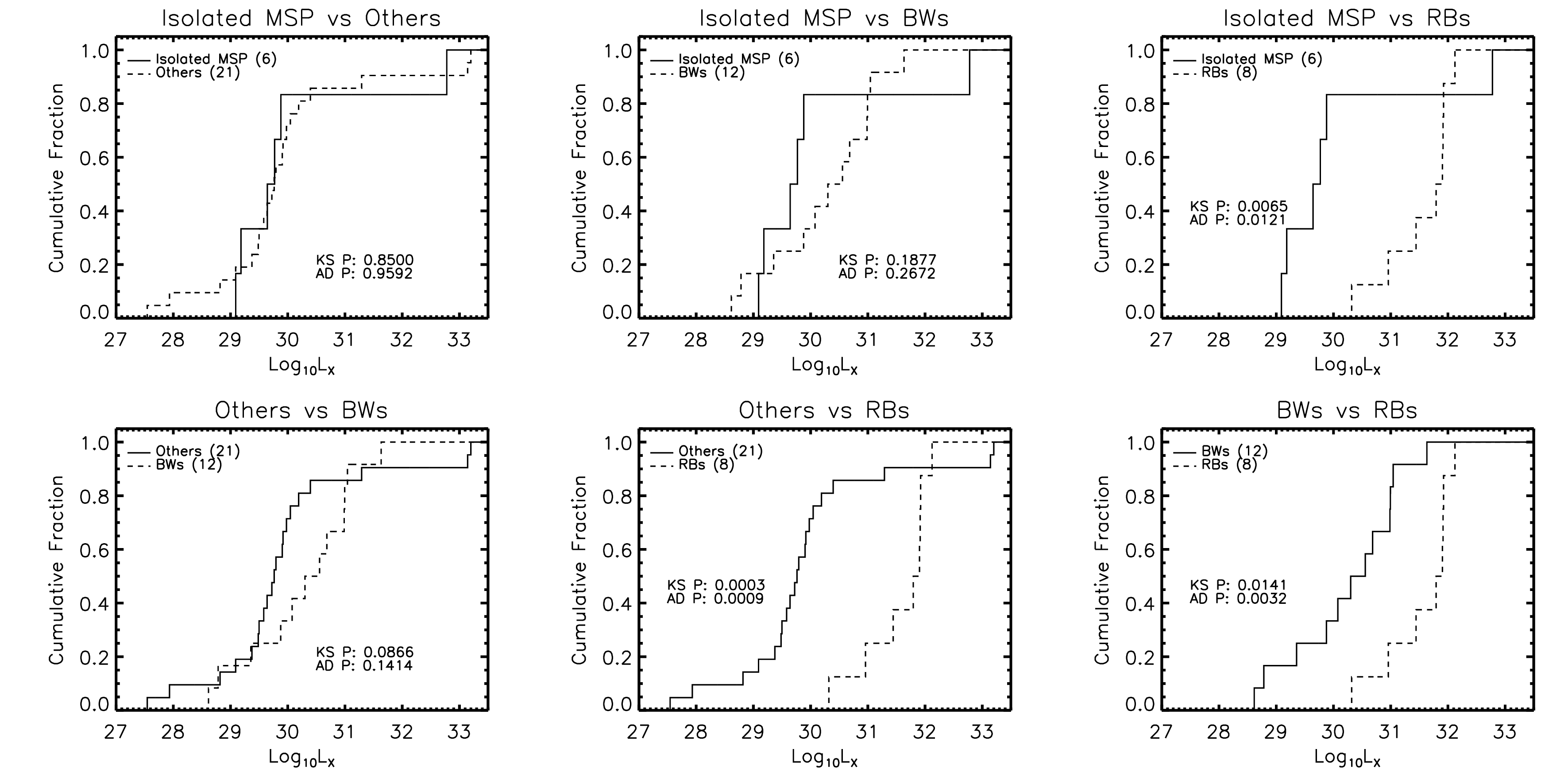}
\caption{Cumulative distribution functions of $L_{x}$ among different classes of X-ray detected MSPs.}
\end{figure}

%\clearpage
\begin{figure}
\centering
\includegraphics[width=
6.0in, angle=0]{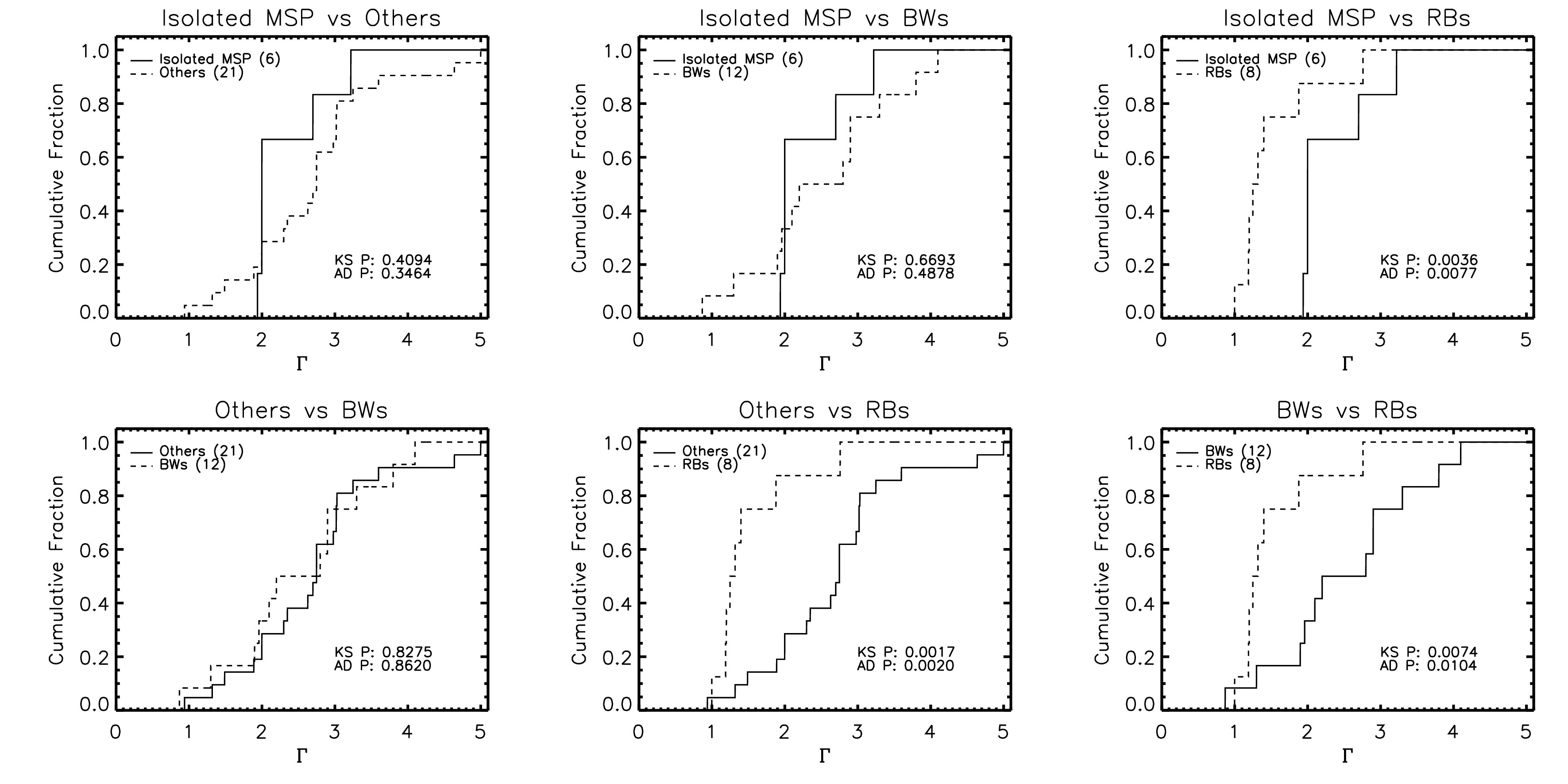}
\caption{Cumulative distribution functions of $\Gamma$ among different classes of X-ray detected MSPs.}
\end{figure}

\clearpage
\begin{figure}
\centering
\includegraphics[width=
6.0in, angle=0]{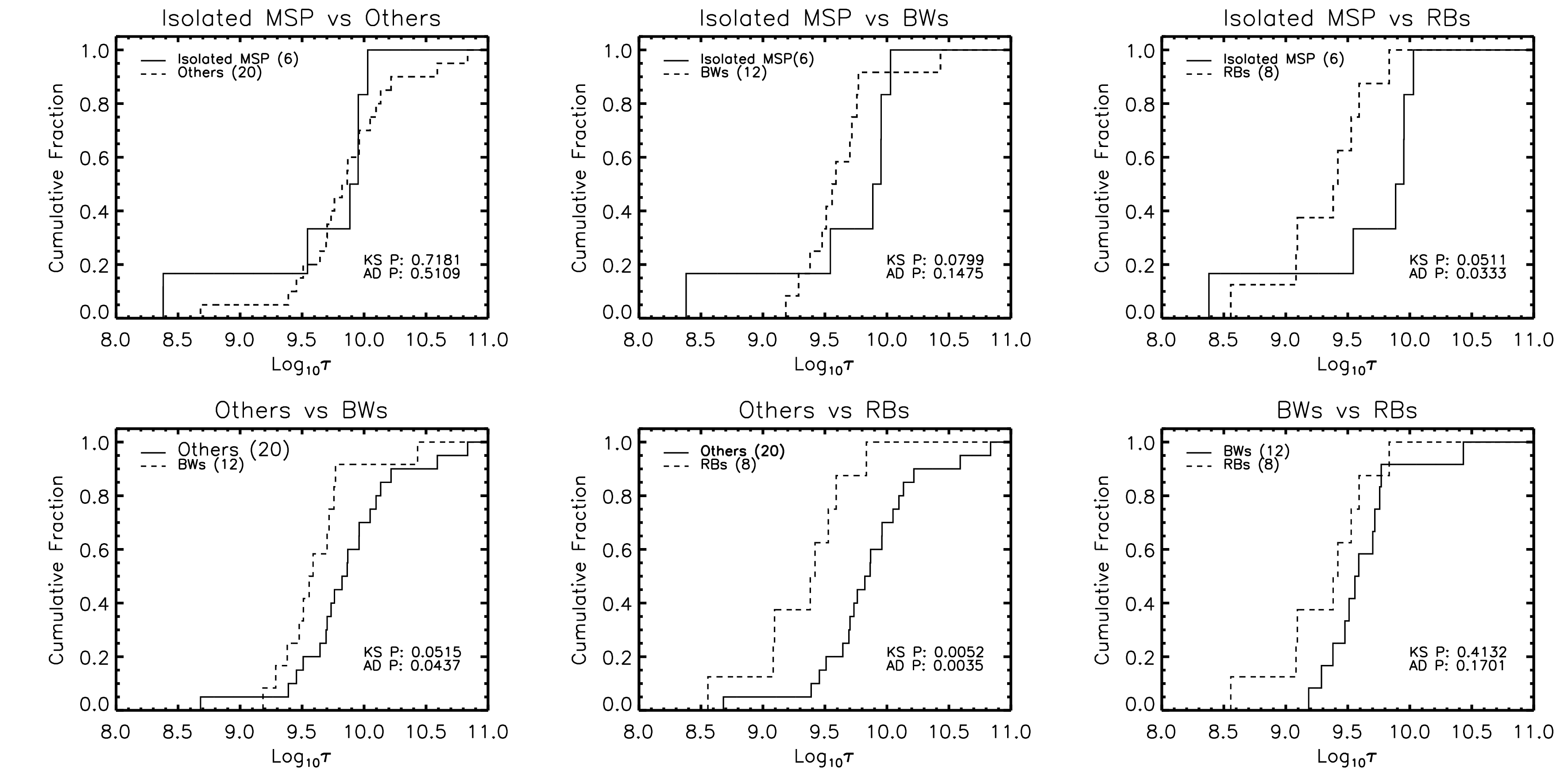}
\caption{Cumulative distribution functions of $\tau$ among different classes of X-ray detected MSPs.}
\end{figure}

%\clearpage
\begin{figure}
\centering
\includegraphics[width=
6.0in, angle=0]{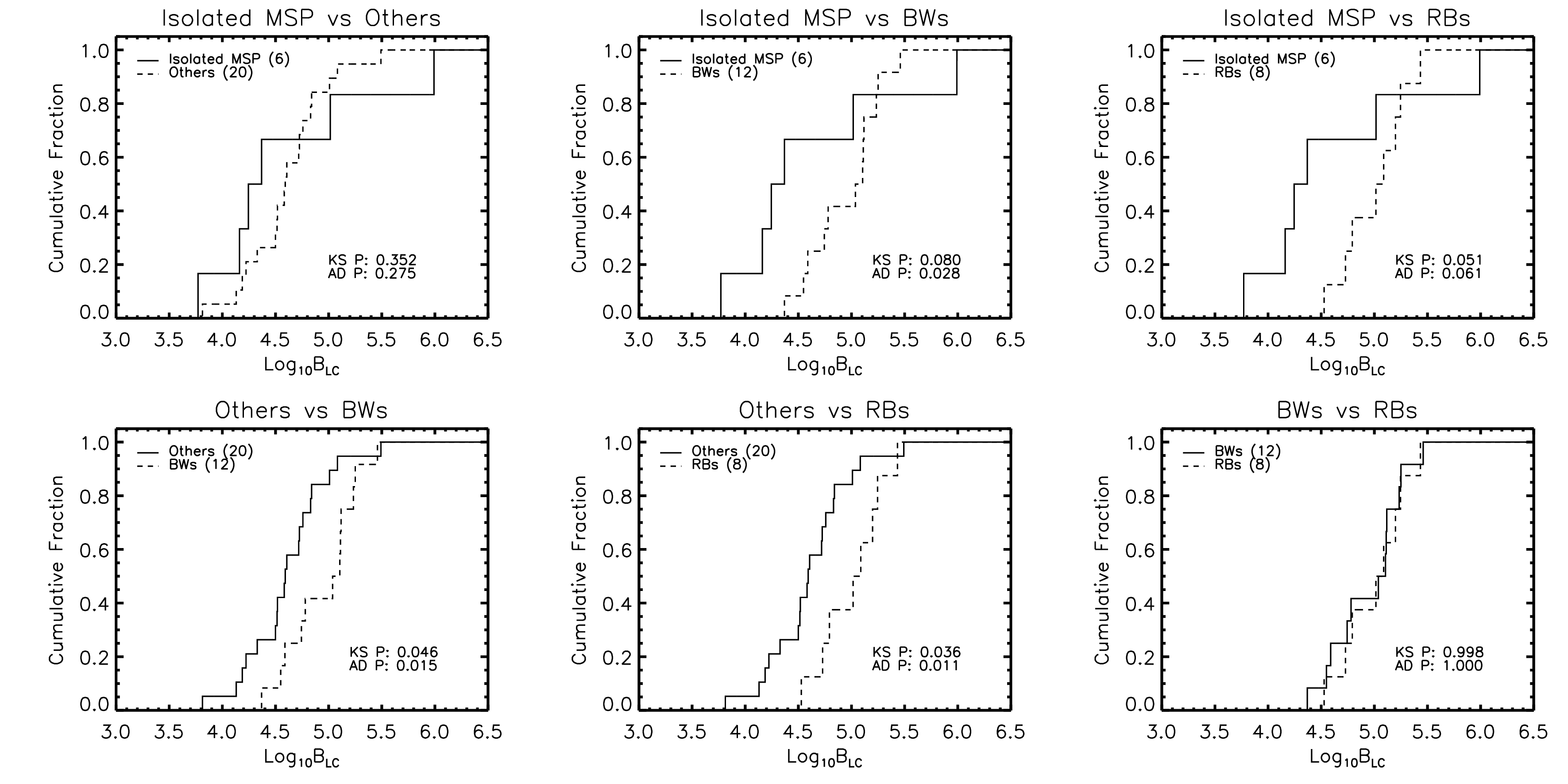}
\caption{Cumulative distribution functions of $B_{LC}$ among different classes of X-ray detected MSPs.}
\end{figure}

\clearpage
\begin{figure}
\centering
\includegraphics[width=
6.0in, angle=0]{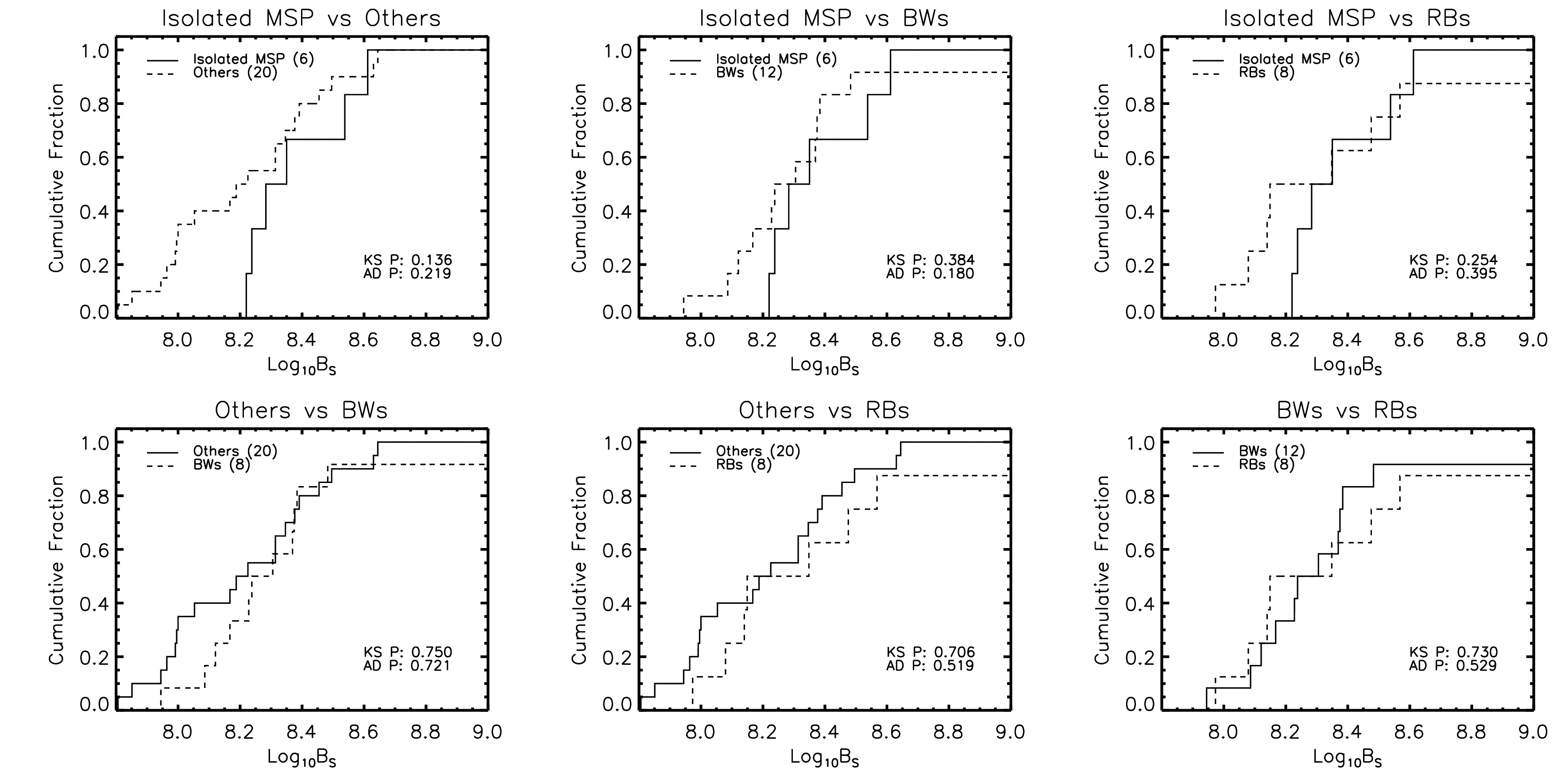}
\caption{Cumulative distribution functions of $B_{S}$ among different classes of X-ray detected MSPs.}
\end{figure}

%\clearpage
\begin{figure}
\centering
\includegraphics[width=
6.0in, angle=0]{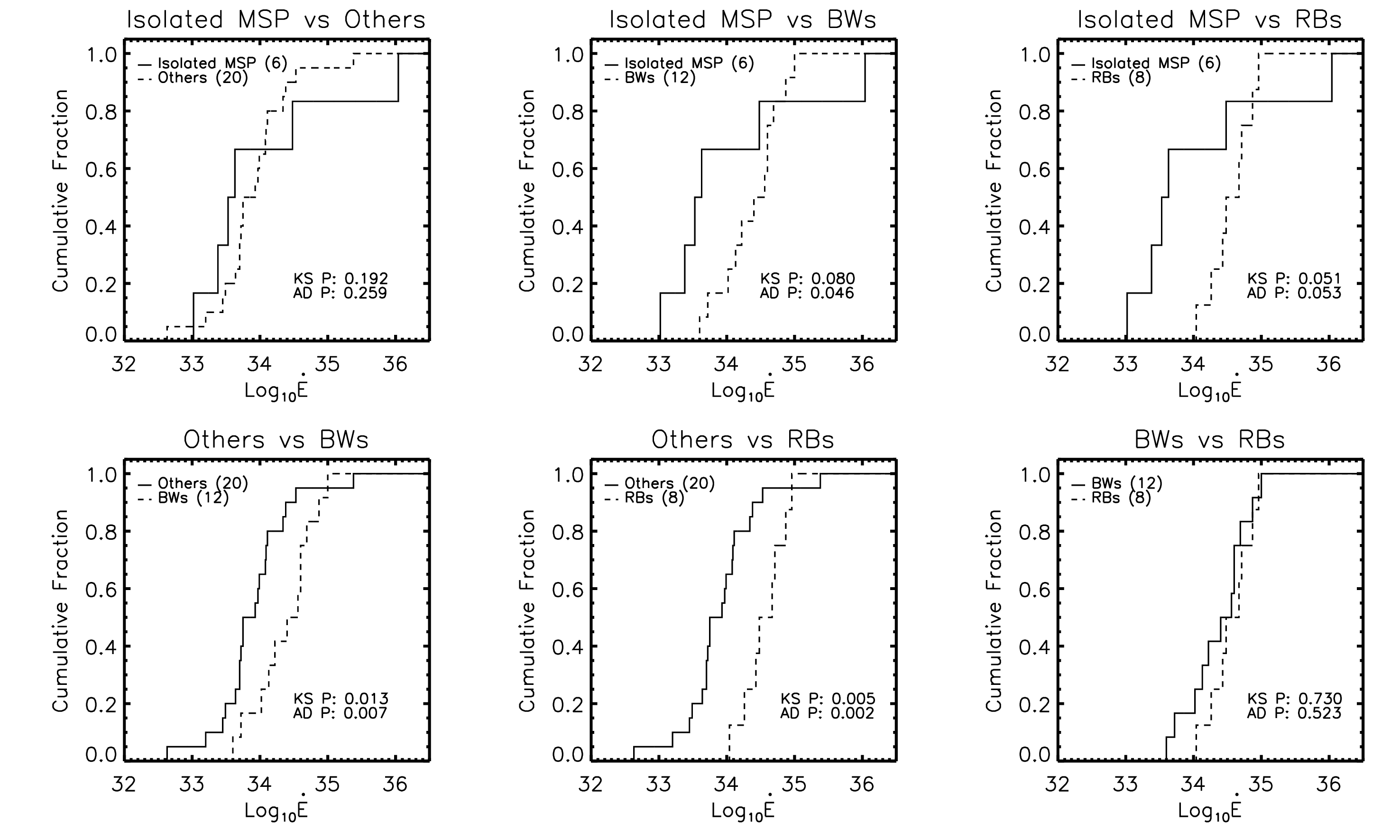}
\caption{Cumulative distribution functions of $\dot{E}$ among different classes of X-ray detected MSPs.}
\end{figure}

\begin{figure}
\centering
\includegraphics[width=
6.0in, angle=0]{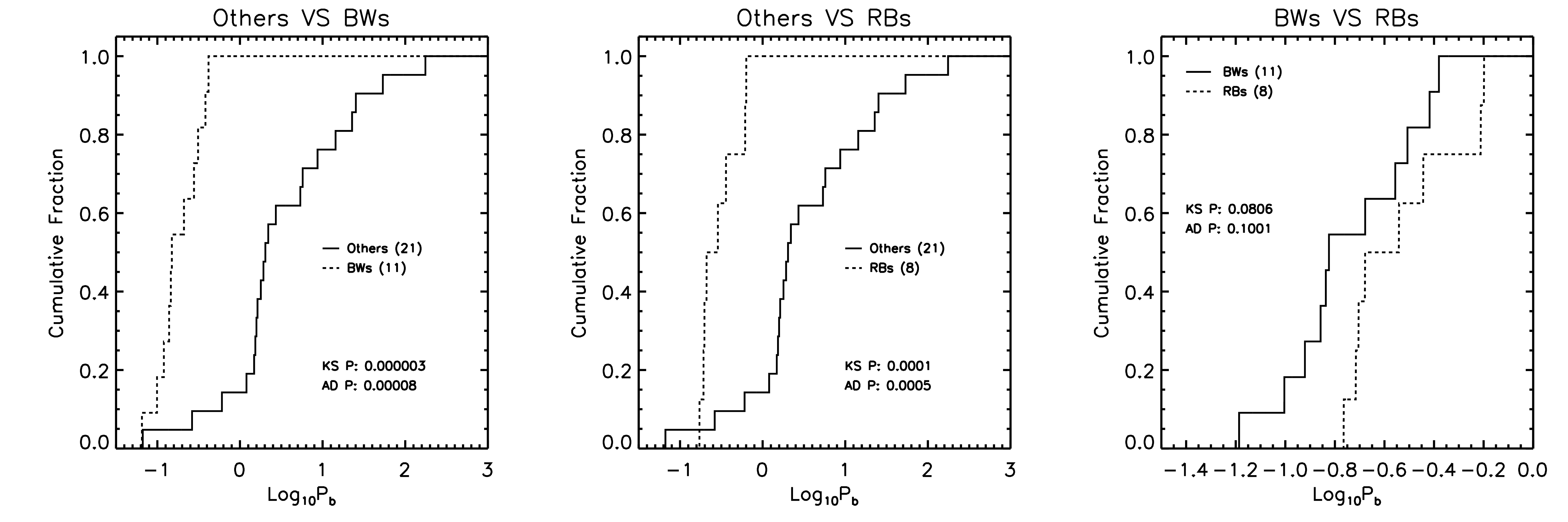}
\caption{Cumulative distribution functions of $P_{b}$ among different classes of X-ray detected MSPs.}
\end{figure}

\begin{figure}
  \centering
  \includegraphics[width=
    6.0in, angle=0]{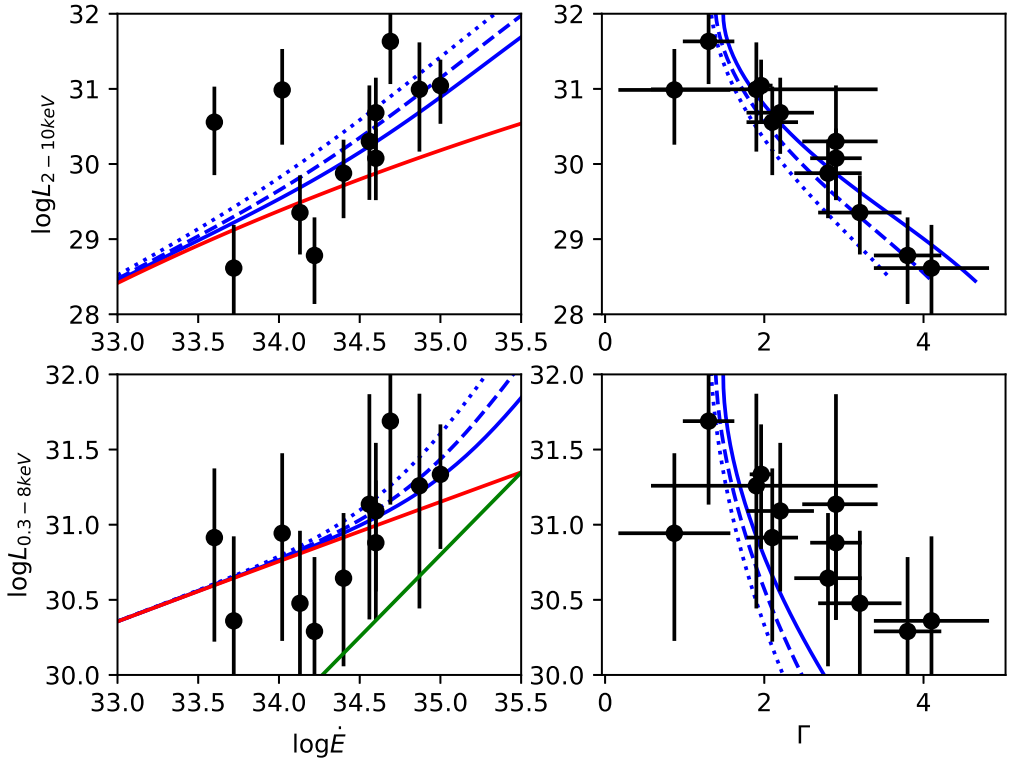}
  \caption{Plots of $L_{x}$ vs $\dot{E}$ ({\it left panels}) and vs $\Gamma$ ({\it right panels}) for the BW MSPs. The top and bottom show the
    X-ray properties in 2-10 keV and in 0.3-8 keV, respectively.
    The green and red lines indicates the magnetospheric synchrotron radiation from
    the secondary pairs [equation (47) in Takata et al. (2012)] and the heated polar cap emission with $B_s=3\time 10^8$Gauss (see text), respectively.
    The blue lines are model prediction with the emission from the intra-binary shock and the heated polar cap.  The solid, dashed and dotted lines are results for
  the efficiency $\delta=0.15$\%, 0.3\% and 0.6\%, respectively.}
\label{bw1}
\end{figure}

\begin{figure}
  \centering
  \includegraphics[width=
    6.0in, angle=0]{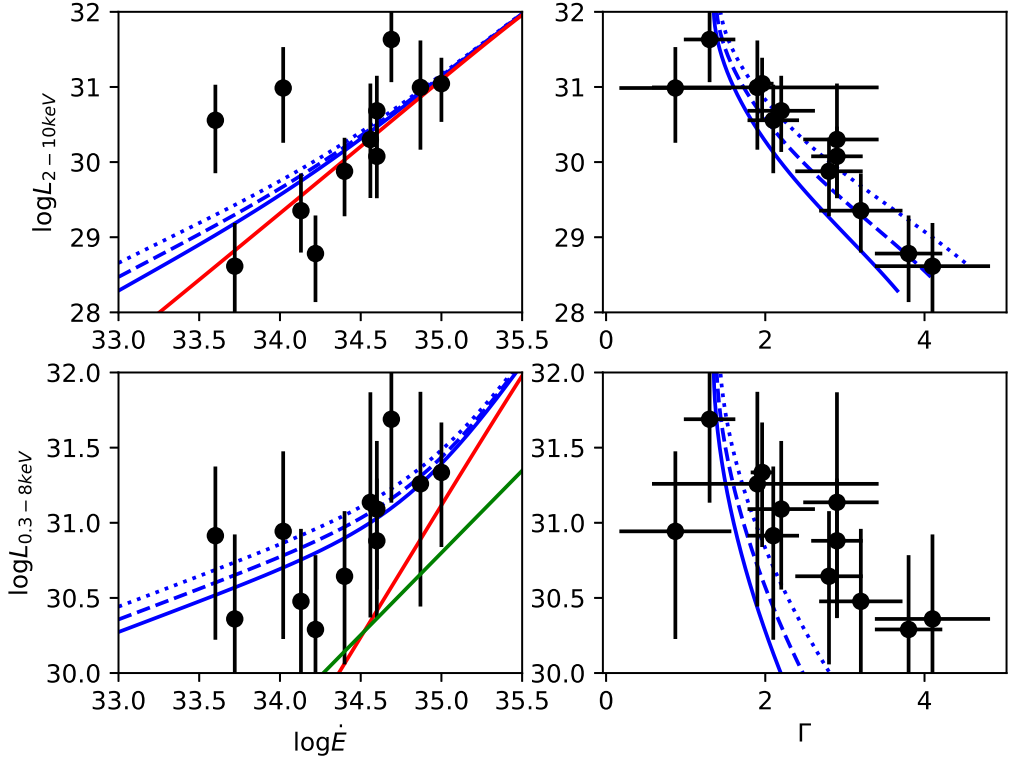}
  \caption{Same as Figure~\ref{bw1}, but the different blue color  lines 
    are results for the different surface magnetic field strength, which affects
    the temperature of the  heated polar cap, as indicated by
    equations~(\ref{tr}) and~(\ref{tc}).The temperatures for the sold, dashed
    and dotted lines are calculated with $B_s=1.5\times 10^8$G, $3\times 10^8$G and
    $6\times 10^8$G, respectively. The red lines
    in the top and bottom panels represent the contribution of the shock emission
    calculated  with $\delta=0.3$\%.}

\label{bw2}
\end{figure}
\begin{figure}
  \centering
  \includegraphics[width=
    6.0in, angle=0]{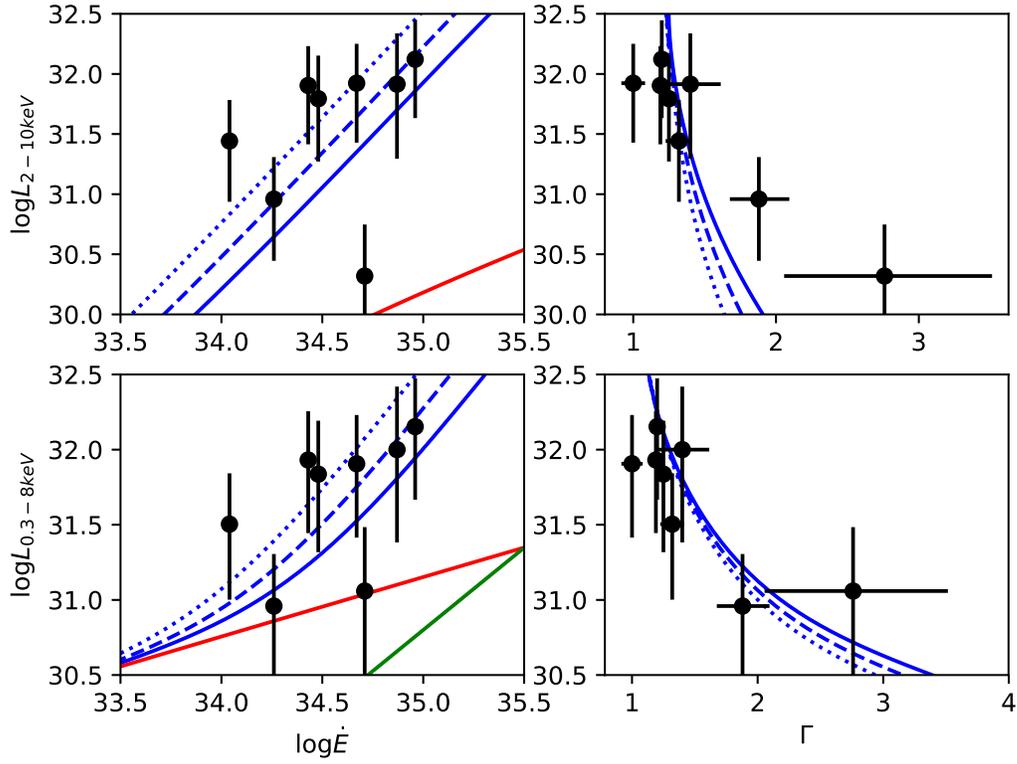}
  \caption{Same as Figure~\ref{bw1}, but for the RB MSPs. The solid, dashed
  and dotted blue lines are results for $\delta=2$\%, 4\% and 8\%, respectively.}
  \label{rb1}
  \end{figure}

\begin{figure}
  \centering
  \includegraphics[width=
    6.0in, angle=0]{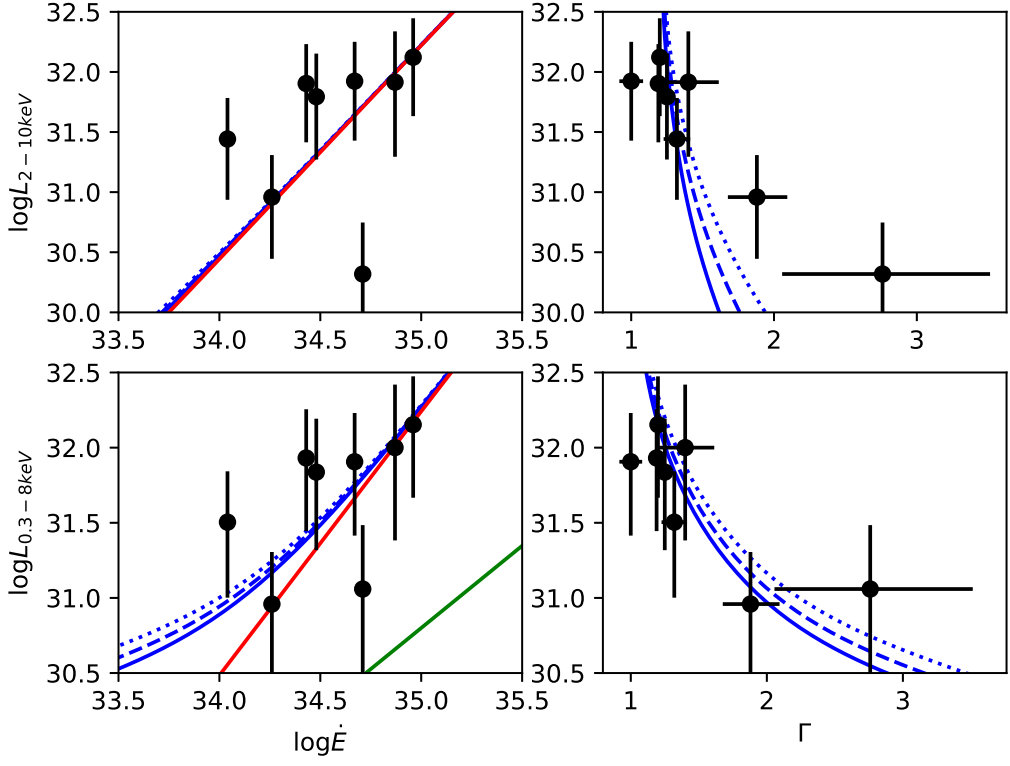}
  \caption{Same as Figure~\ref{bw2}, but for the RB MSPs. The contribution of
  the shocked emission (red lines) is calculated with $\delta=4\%$.}
  \label{rb2}
    \end{figure}

%\clearpage

\end{document}